\documentclass[aps,pra,amsmath,amssymb,nofootinbib]{revtex4} 

\usepackage{color}
\usepackage{textcomp}
\usepackage{amsmath}
\usepackage{amssymb}
\usepackage{graphicx}
\usepackage{dcolumn}
\usepackage{bm}
\usepackage{bbold}
\usepackage{pslatex}
\usepackage{epsfig}
\usepackage{color}
\usepackage{dsfont}
\usepackage[unicode=true,pdfusetitle,
 bookmarks=true,bookmarksnumbered=false,bookmarksopen=false,
 breaklinks=false,pdfborder={0 0 1},backref=false,colorlinks=true]
 {hyperref}
\usepackage{breakurl}
\usepackage[normalem]{ulem} 

\newcommand{\ket}[1]{\ensuremath{|#1\rangle}}
\newcommand{\bra}[1]{\ensuremath{\langle #1|}}

\newcommand{\be}{\begin{equation}}
\newcommand{\ee}{\end{equation}}
\newcommand{\ba}{\begin{eqnarray}}
\newcommand{\ea}{\end{eqnarray}}

\newcommand{\rchm}[2]{#1} 

\begin{document}


\title{High Resolution non-Markovianity in NMR}

\author{Nadja K. Bernardes$^{1,*}$, John P. S. Peterson$^{2}$, Roberto S. Sarthour$^{2}$, Alexandre M. Souza$^{2}$, C. H. Monken$^{1}$, Itzhak Roditi$^{2}$, Ivan S. Oliveira$^{2}$, Marcelo F. Santos$^{1}$}

\affiliation{$^{1}$Departamento de F\'isica, Universidade Federal de Minas Gerais, Belo Horizonte, Caixa Postal 702, 30161-970, Brazil}

\affiliation{$^{2}$Centro Brasileiro de Pesquisas F\'isicas, Rua Dr.  Xavier Sigaud 150, Rio de Janeiro, 22290-180, Brazil}
\address{$^*$nadjakb@fisica.ufmg.br}






\date{\today}


\begin{abstract}
Memoryless time evolutions are ubiquitous in nature but often correspond to a resolution-induced approximation, i.e. there are correlations in time whose effects are undetectable. Recent advances in the dynamical control of small quantum systems provide the ideal scenario to probe some of these effects. Here we experimentally demonstrate the precise induction of memory effects on the evolution of a quantum coin (qubit) by correlations engineered in its environment. In particular, we design a collisional model in Nuclear Magnetic Resonance (NMR) and precisely control the strength of the effects by changing the degree of correlation in the environment and its time of interaction with the qubit. 
We also show how these effects can be {\it hidden} by the limited resolution of the measurements performed on the qubit. The experiment reinforces NMR as a test bed for the study of open quantum systems and the simulation of their classical counterparts. 
\end{abstract}

\maketitle

\section*{Introduction} 
Markov processes are defined as those in which the future of a system bears no correlation to its past. Such behavior is found in the description of phenomena as far apart as nuclear fission, the decay of an electronically excited atom, the growth of bacteria colonies or rabbit populations and the compound interests that dictate our debts, to name a few~\cite{kampen}. In many situations, however, these time evolutions prove to be no more than a resolution-induced approximation, i.e.  the interaction with external degrees of freedom (here called environment) produces correlations in the dynamics, but in a way that is too weak to be observed. Most of the times, such small deviations are either harmless and/or useless. However, there are meaningful counterexamples of both cases. For instance, many modern applications, such as the generation of large cryptographic keys or statistical sampling, rely on random number generators~\cite{chaitin}. Testing the independence of events in these devices is essential to guarantee the safety or fairness of these applications where even the tinniest correlations may be exploited. At the same time, controllable non-Markovian evolutions in small quantum systems have been thought of as promising tools for quantum information processing~\cite{vasile,matsuzaki,chin,elsi,Bogna}.

Markovian processes exist both in discrete time steps or as a continuous time evolution. A broadly used tool to discretize continuous Markovian dynamics are the so-called collisional models. In these models, the continuous stochastic process is replaced by a stroboscopic sequence formed by a series of discrete time steps defined, each one, by a collision between the system of interest and one of the particles forming its environment. The most famous application of collisional models in physics is the description of Brownian motion. In open quantum systems, collisional models have been extensively used as a test bed to investigate the details of system-environment interactions and the conditions under which non-Markovian dynamics arises~\cite{rau,ziman1,ziman2,giovannetti,tomas,ciccarello,vacchini,budini,Paternostro,nadja1}. Provided system and environment are initially uncorrelated, the evolution of the system $\rho$ after $j$ collisions with the environment $\rho_{env}$ is given by
\be\label{rhof}
\rho\rightarrow\rho'=\text{Tr}_{env}\left[U_jU_{j-1}...U_2U_1(\rho\otimes \rho_{env})U_1^{\dagger}U_2^{\dagger}...U_{j-1}^{\dagger}U_j^{\dagger}\right],
\ee
where $\text{Tr}_{env}$ denotes the partial trace over the environmental degrees of freedom and $U_i$ is a global unitary operation due to the $i$-th collision. These models are simple to describe, easy to compute and have been recently simulated in optical setups where one degree of freedom of the photon is used as the quantum system and extra ones play the role of the environment~\cite{Liu,Liu2,tang,Steve1,Fabio2,chiuri,Xu,Fabio1,nadja2}.  

In this paper, we implement a collisional model in Nuclear Magnetic Resonance (NMR) to investigate with high resolution the subtle non-Markovian evolution of a qubit subjected to the action of a structured and fully quantized external environment. This experimental technique has been very successful in the study of quantum information processes and in proofs of quantum principles as may be seen on \cite{t1,t2,t3,t4}. Here, the environment is composed of independent qubits that are initially correlated among themselves and collide, one at a time, with the target. We characterize non-Markovianity by calculating the Trace Distance \rchm{$D(\rho_1(t),\rho_2(t))= \frac{1}{2}\text{Tr}|\rho_1(t)-\rho_2(t)|$}{$D(t)= \frac{1}{2}\text{Tr}|\rho_1(t)-\rho_2(t)|$} between two different states of the qubit as a function of time. It is well established that in a quantum stochastic evolution, \rchm{$D(\rho_1(t+\tau),\rho_2(t+\tau))>D(\rho_1(t),\rho_2(t))$}{$D(t+\tau)>D(t)$} implies non-Markovian behavior~\cite{BLP}. The experiment features only two time steps because these are enough to establish all the properties we want to discuss. We measure \rchm{$D(\rho_1(t),\rho_2(t))$}{$D(t)$} after each collision as a function of the degree of correlation in the initial state of the environment and also as a function of the duration of each collision. Both quantities influence how much the dynamics deviates from a Markov process. 

\section*{Results}
\begin{figure}[t!]
\centering
\includegraphics[scale=1.2]{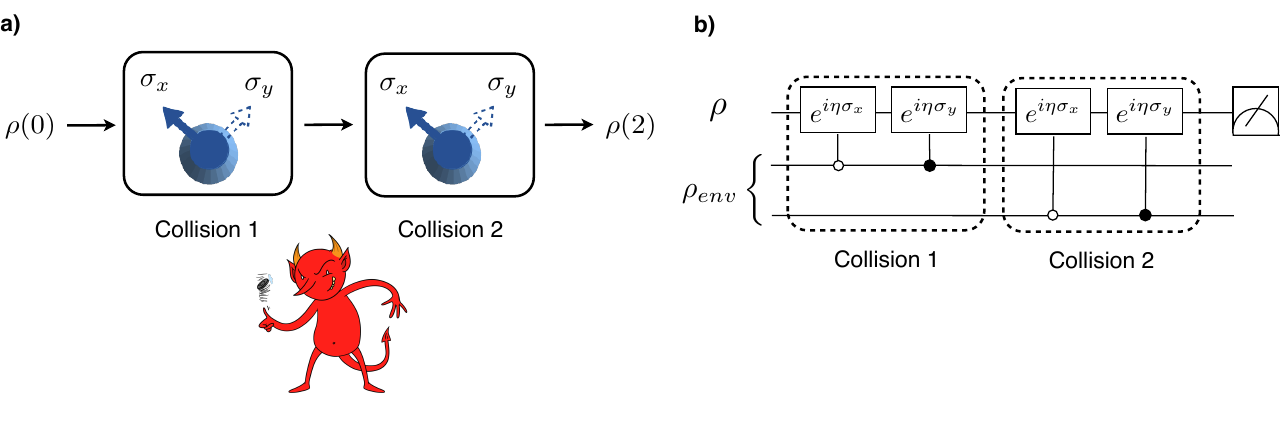}
\caption{a) The nuclear spins of the atoms of fluorine of the C$_{2}$F$_{3}$I molecule are the qubits. Two of them represent the environment and the other one is the system. Their interaction is showed in this pictorial representation of our collisional model. The action of the environment on the system's qubit is represented by the two boxes and the demon. In the first collision, the demon will toss a coin and will decide if he will operate either $e^{i\eta\sigma_x}$ or $e^{i\eta\sigma_y}$ on the system $\rho(0)$. Afterwards he can still decide if the operation done in the second collision is correlated or not with the previous one, resulting in the state $\rho(2)$. b) A quantum circuit that describes the experiment. The environment is prepared in state $\rho_{env}$ and will serve as the control qubit for the controlled operation that happen on the state of the system $\rho$. After that a quantum state tomography is performed on the first particle in order to determine the state of the system.}
\label{circuit}
\end{figure}

We encoded the system-environment state in a sample of trifluoroiodoethylene (C$_{2}$F$_{3}$I). The NMR experiment was performed using a 500 MHz Varian spectrometer with the used sample diluted ($\sim 1\%$) in deuterated acetone (containing $97\%$ of deuterium). A single molecule of C$_{2}$F$_{3}$I contains three atoms of fluorine with nuclear spin-$1/2$, each of them representing a qubit. One of the atoms will be regarded as the system and the other two as the environment. These spins interact with each other and with applied magnetic fields as well. We designed the experiment to implement the operations represented in the quantum circuit of Fig.~\ref{circuit}: the first qubit, top line in the circuit, represents the system and the other lines are the qubits of the environment. The two qubits of the environment are initially prepared in a correlated state
$\rho_{env} = \frac{q}{2}(\ket{00}\bra{00}+|11\rangle\langle11|)+\frac{1-q}{2}(|01\rangle\langle01|+|10\rangle\langle10|)$\, where ``0'' means spin up and ``1'' means spin down. In each collision, the state of the system undergoes a random-walk type of evolution, changing conditionally to the environment: environmental spin up induces a rotation in a certain direction (say, $x$), whereas environmental spin down induces a rotation in the orthogonal direction ($y$). This situation can be summed up by the unitary transformation $U=e^{i\eta\sigma_x}\otimes |0\rangle\langle 0| + e^{i\eta\sigma_y}\otimes |1\rangle\langle 1|$ where $|j\rangle\langle j|$ represents the internal state of an environmental particle and the exponentials dictate the system's rotations.

In a NMR experiment
the signal of each fluorine spin can be distinguished and singled
manipulated. The natural Hamiltonian of the total system may be described as
\begin{equation}
\mathcal{H}=\sum_{n}\hslash\left(\omega_{0n}-\omega_{r}\right)I_{z}^n+\sum_{k\neq m}\hslash2\pi\mathcal{J}_{km}I_{z}^k\otimes I_{z}^m,
\label{eq:nh}
\end{equation}
where $\omega_{0n}$
and $\omega_{r}$ are the natural resonance frequency and the frequency
of the rotating reference frame of the $n$-th spin, respectively. The first term represents the interaction of the spins with the static
magnetic field applied along the $z$ direction and the second
term represents the interaction between the three spins, with $\mathcal{J}_{km}$ being the exchange integral. Here, $I_{\alpha}^i$ is the spin operator for the $i$-th spin, which for spin-$1/2$ corresponds to the Pauli matrix ($\sigma_{\alpha}$)
divided by 2. The physical parameters of our molecule are given in Methods. Since the individual chemical shifts of each spin place them far
apart, i.e. their resonance frequencies are separated enough ($\left|\omega_{0k}-\omega_{0m}\right|\gg\mathcal{J}_{km}$),
the Ising coupling approximation was considered here \cite{key-2}. In the
NMR experiment the spins also interact with oscillating magnetic fields
that can be turned on and off at any frequency and different amplitudes,
named radio frequency pulses ($rf$). These are responsible for the
system control, and its interaction can de described by the Hamiltonian 
\begin{equation}
\mathcal{H}_{rf}=\hslash\omega_{1}(t)\left[I_{x}^n\cos\phi +I_{y}^n\sin\phi \right],\label{eq:rf}
\end{equation}
where $\omega_{1}(t)$
defines the pulse modulation and duration, and $\phi$ stands for
its phase. The radio frequency pulses are used to perform rotations on the individual
spins and by properly choosing these rotations, the quantum circuit from Fig.~\ref{circuit} can be implemented, as described in Methods. 

The experiment was performed for two different initial states of the system, $\rho_1(0)=\ket{0}\bra{0}$ and $\rho_2(0)=\ket{1}\bra{1}$ (eigenstates of
$\sigma_z$) and for the environment prepared in different states with different degrees of correlation $q$ ($q=0$, 0.15, and 0.25). The theoretical analysis shows that for $q$ equals $0$ and $0.15$ these are the states that maximize $\Delta D$ for the chosen dynamics. For $q=0.25$ there is a better pair of initial states but the eventual gain in $\Delta D$ (a factor of $\sim 3$) still lies within the current error bar. Using the $rf$ pulses, the first and the second collisions were implemented, resulting in the system state $\rho_1(1)$ ($\rho_2(1)$) and $\rho_1(2)$ ($\rho_2(2)$) , respectively. After each run, full state tomography~\cite{key-1} was performed in the system.

\begin{figure}[t!]
\centering
\includegraphics[scale=1.3]{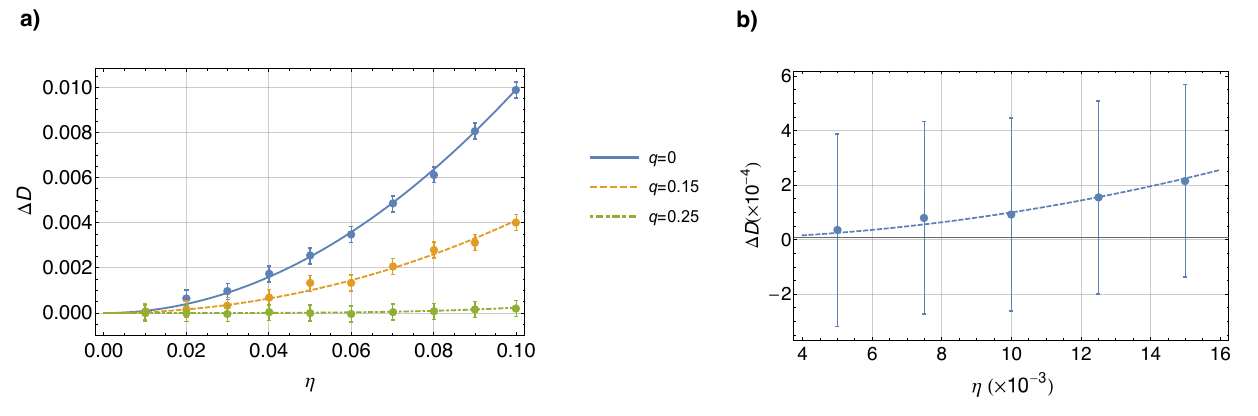}
\caption{The trace distance, $D(\rho_1(t),\rho_2(t))= \frac{1}{2}\text{Tr}|\rho_1(t)-\rho_2(t)|$, was employed to witness non-Markovianity. The increasing of $D(\rho_1(t),\rho_2(t))$ with time is a trace of non-Markovianity. a) The plots show the dependence of $\Delta D$ with the strength of each collision $\eta$ for different values of the correlation parameter {$q$ (0 for the solid blue line, 0.15 for the dashed orange line and 0.25 for the dash-dotted green line). $q=0$ means that the two collisions were totally anti-correlated. b) Detail of the $\Delta D$ curve for $\eta<0.01$ and $q=0$. In both plots the error bars were estimated taking into account every possible error source, as discussed in Ref.~\cite{cesar}. This led to the error determining the Bloch vector of the quantum state of $\delta r=5\times 10^{-4}$, independent of the direction. By propagating this error, the error in $D(\rho_1(t),\rho_2(t))$} is given by $\delta D=\delta r/\sqrt{2}$.}
\label{fig1}
\end{figure}

Figure~\ref{fig1} shows the change in distance between the two initial states of the system after one and two collisions $\Delta D = D(\rho_1(2),\rho_2(2))-D(\rho_1(1),\rho_2(1))$,
as a function of the strength of each collision $\eta$ and for different degrees of correlation $q$ of the environmental state. For large enough interactions (larger $\eta$) and anti-correlation in the environment (smaller $q$), the collisions clearly generate a non-Markovian dynamics in the system ($\Delta D>0$). The phenomenon, witnessed by the increase in $\Delta D$ also reflects a backflow of information from the environment to the system as time progresses. 

It is also worth noticing that, as the non-Markovian effects induced by the collisions get smaller, either due to weaker collisions (small values of $\eta$) and/or environmental correlations ($q \rightarrow 1/2$), the uncertainties in $\Delta D$ due to the lack of resolution of the measurements become of the same order of $\Delta D$.
This is true even for the strongest anti-correlation possible, $q=0$. In this case, $\Delta D$, a sufficient but not necessary condition for non-Markovianity, does not provide 
a conclusive answer anymore, even though theory still predicts its vanishing only asymptotically with both parameters $\eta\rightarrow 0$ or $q\rightarrow 1/2$. Our resolution to conclusively guarantee non-Markovianity was limited to $\Delta D \sim 3.5\times10^{-4}$.

\section*{Discussions}
Non-markovian dynamics of collisional models has been experimentally investigated in linear optical setups~\cite{Liu,Liu2,tang,Steve1,Fabio2,chiuri,Xu,Fabio1,nadja2} where the environment is simulated by semi-classical degrees of freedom. Usually, such experiments are performed in the photon counting regime and strong system-environment coupling. Our results put in evidence, in a controllable way, the role played by environmental correlations in the non-Markovian dynamics of quantum systems and also show how such effects may be hidden by the external noise brought in by unavoidable measurement uncertainties. We have shown how NMR can be exploited to study quantum dynamics of open systems using a bona fide quantum environment. NMR also presents a great advantage over previously used platforms when it comes to the versatility to design and control different environmental states and the precision to probe very small memory effects in the dynamics of quantum systems. For example, here we have observed variations of the Trace Distance of the order of $0.05\%$, a gain of at least two orders of magnitude over the typical optical experiments based on measurements of the polarization of photons. In photon polarization measurements, the signal-to-noise ratio is less than $\sqrt{N}$, where $N$ is the number of photons in each measurement. Such gain in precision can be essential, for example, if one would like to use a similar setup to test very small time correlations of black boxes such as quantum or classical random number generations, cryptographic machines and alike. Furthermore, one can also explore the capacity to dynamically change the environment during the interaction with the system and to directly interfere with the system during its time evolution in order to test and investigate very recent theoretical results in the field, such as the ones respectively published in ~\cite{ciccarello} and ~\cite{simone}.

\section*{Methods}
\subsection*{Model}
We consider a qubit $\rho$ that interacts with an environment from which it is initially decoupled. The interaction consists of consecutive collisions with qubits of the environment. During each collision, the system undergoes with equal probability an evolution $e^{i\eta\sigma_y}$ or $e^{i\eta\sigma_x}$ controlled by the internal state of the corresponding environmental particle. The map that describes the state of the system after collision one is $\Lambda_{10}(\cdot)=\frac{1}{2}[e^{i\eta_1\sigma_x}(\cdot)e^{-i\eta\sigma_x}+e^{i\eta_1\sigma_y}(\cdot)e^{-i\eta_1\sigma_y}]$.
All the effects we are interested in can be observed with only two collisions described, in our case, by the general map $\Lambda_{20}(\cdot)=\frac{q}{2}[e^{i\eta_2\sigma_x}e^{i\eta_1\sigma_x}(\cdot)e^{-i\eta_1\sigma_x}e^{-i\eta_2\sigma_x}+e^{i\eta_2\sigma_y}e^{i\eta_1\sigma_y}(\cdot)e^{-i\eta_1\sigma_y}e^{-i\eta_2\sigma_y}]+\frac{1-q}{2}[e^{i\eta_2\sigma_y}e^{i\eta_1\sigma_x}(\cdot)e^{-i\eta_1\sigma_x}e^{-i\eta_2\sigma_y}+e^{i\eta_2\sigma_x}e^{i\eta_1\sigma_y}(\cdot)e^{-i\eta_1\sigma_y}e^{-i\eta_2\sigma_x}]]$, where $q$ defines their degree of correlation: for $q=1$ ($q=0$) the collisions are totally correlated (anti-correlated). One possible way to generate this evolution is to consider two qubits as the environment prepared in the following state $\rho_{env}=\frac{q}{2}(\ket{00}\bra{00}+\ket{11}\bra{11})+\frac{1-q}{2}(\ket{01}\bra{01}+\ket{10}\bra{10})$. (Note that a classically correlated state or the state $\sqrt{\frac{q}{2}}(\ket{00}+\ket{11})+\sqrt{\frac{1-q}{2}}(\ket{01}+\ket{10})$ will render the same system evolution). The interaction in each collision is described by a unitary transformation given by $U_{\eta}=e^{i\eta\sigma_x}\otimes\ket{0}\bra{0}+e^{i\eta\sigma_y}\otimes\ket{1}\bra{1}$, which works as a conditional operation. The resulting state after one collision will be $\rho(1)=\Lambda_{10}(\rho(0))=\text{Tr}_{env}\left[(U_{\eta_1}\otimes \mathds{1})(\rho(0)\otimes\omega_{env})(U_{\eta_1}^{\dagger}\otimes \mathds{1})\right]=
\frac{1}{2}(e^{i\eta_1\sigma_x}\rho(0)e^{-i\eta_1\sigma_x}+e^{i\eta_1\sigma_y}\rho(0)e^{-i\eta_1\sigma_y})$, where $\text{Tr}_{env}$ denotes the partial trace over the environmental degrees of freedom. And after the second collision the evolution of the system reads $\rho(2)=\Lambda_{20}(\rho(0))=\text{Tr}_{env}\left[U_{\eta_2}U_{\eta_1}(\rho(0)\otimes\omega_{env})U^{\dagger}_{\eta_1}U^{\dagger}_{\eta_2}\right]=\frac{q}{2}[e^{i\eta_2\sigma_x}e^{i\eta_1\sigma_x}(\rho(0))e^{-i\eta_1\sigma_x}e^{-i\eta_2\sigma_x}+e^{i\eta_2\sigma_y}e^{i\eta_1\sigma_y}(\rho(0))e^{-i\eta_1\sigma_y}e^{-i\eta_2\sigma_y}]+\frac{1-q}{2}[e^{i\eta_2\sigma_y}e^{i\eta_1\sigma_x}(\rho(0))e^{-i\eta_1\sigma_x}e^{-i\eta_2\sigma_y}+e^{i\eta_2\sigma_x}e^{i\eta_1\sigma_y}(\rho(0))e^{-i\eta_1\sigma_y}e^{-i\eta_2\sigma_x}]]$. For simplicity, we chose these two specific operators, however, similar results would be observed had we used two different non-commuting operators. The parameters $\eta_1$ and $\eta_2$ are related to the strength of interaction and the time for one collision.  We consider $\eta_1=\eta_2=\eta$, and considering a weak interaction and a short collision time, $\eta$ are assumed to be sufficiently small ($\eta<0.1$) such that $e^{i\eta\sigma_{x,y}} \sim(1+i\eta\sigma_{x,y})$. 

In the first run of the experiment, the system interacts (collides) with qubit one of the environment and, in the second run, 
the system repeats this collision and subsequently collides with qubit two of the environment. 
After each run, quantum state tomography \cite{key-1} is performed on the system in order to fully determine its state. Two orthogonal 
initial states of the system, $\rho_{1}(0)=\left|0\right\rangle \left\langle 0\right|$ and $\rho_{2}(0)=\left|1\right\rangle \left\langle 1\right|$,
are prepared so that the distance $D(\rho_1(j),\rho_2(j))$ between them can be calculated after each run $j$ (orthogonality sets $D(\rho_1(0),\rho_2(0))=1$).

Being a qubit, the state of the the system can also be represented by a vector $\vec{r}=(\langle \sigma_x\rangle,\langle \sigma_y\rangle,\langle \sigma_z\rangle)$ in the Bloch Sphere ($\langle \sigma_i\rangle = \textrm{Tr}\{\sigma_i \rho\}$) in which case the distance between $\rho_1$ and $\rho_2$ reduces to the geometric distance between the respective vectors $D(\rho_1,\rho_2)=\frac{1}{2}|\vec{r}_1-\vec{r}_2|$. This quantity can be easily calculated for the initial states $\rho_1(0) = |0\rangle\langle0|$ and $\rho_2(0) = |1\rangle\langle1|$ and for the rotations used in the experiment and gives $D(\rho_1(1),\rho_2(1))=\frac{1}{2}[3 + \cos(4\eta)]^\frac{1}{2}$ and $D(\rho_1(2),\rho_2(2))=\frac{1}{2} \left\{16[\cos\eta \sin\eta(\cos ^2\eta -q \sin^2\eta)]^2+[(q+1) \cos (4\eta )-q+1]^2+4 \sin^2\eta\cos^2\eta [(q+1) \cos (2\eta)-q+1]^2\right\}^\frac{1}{2}$ after the first and second collision, respectively. Expanding these results for $\eta \ll 1$ we get $\Delta D = (1-4q) \eta^2 + O(\eta^4)$, making it clear why, for the range of $\eta$ used in the experiment, $\Delta D$ must be positive, i.e. the dynamics must be non-Markovian, for any $q<1/4$ (highly anti-correlated environmental qubits). This result corresponds to the solid curves in Fig.~\ref{fig1}.

\subsection*{Experimental details}
\begin{figure}
\includegraphics[scale=1.2]{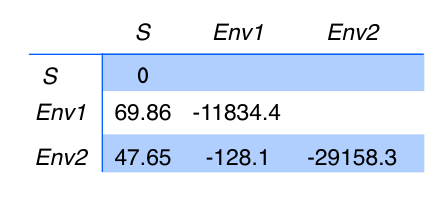}
\caption{Table of $(\omega_{0k}-\omega_r)/(2\pi)$ in the main
diagonal elements and $\mathcal{J}_{km}$ as the other elements,
all values are given in Hz.}
\label{tabelaa}
\end{figure}

The NMR experiment was performed encoding the system of interest and environment in a trifluoroiodoethylene (C$_{2}$F$_{3}$I) sample diluted ($1\%$) deuterated acetone (containing $97\%$ of deuterium). The three nuclei spin-$1/2$ of the fluorine atoms were associated to system and environment qubits. The total system is described by the Hamiltonian
\begin{equation}
\mathcal{H}=\sum_{n}\hslash\left(\omega_{0n}-\omega_{r}\right)I_{z}^n+\sum_{k\neq m}\hslash2\pi\mathcal{J}_{km}I_{z}^k\otimes I_{z}^m,
\label{eq:nh}
\end{equation}
where $\omega_{0n}$ and $\omega_{r}$ are the natural resonance frequency and the frequency
of the rotating reference frame of the $n$-th spin, $I_{z}^n$ is the $z$-component of the spin angular moment of nucleus $n$. The second
term represents the interaction - being $\mathcal{J}_{km}$ the exchange
integral - between the three spins.  Fig.~\ref{tabelaa} shows the values of $(\omega{}_{0n}-\omega_{r})$
and $\mathcal{J}_{km}$. 

\subsection*{Preparation of the initial state and unitary operations}
\begin{figure}[t!]
\centering
\includegraphics[scale=1.4]{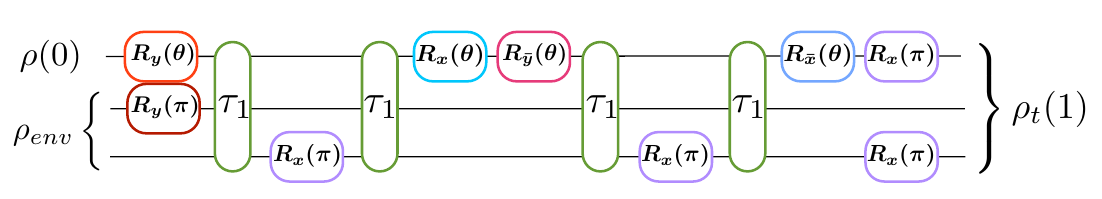}
\caption{Quantum circuit representing the NMR pulse sequence for implementing the first collision. The boxes with the symbols $R_{\alpha}(\theta)$
indicate that a rotation of the angle $\theta$ on that particular
spin was performed around the $\alpha$ direction and $R_{\bar{\alpha}}(\theta)=R_{\alpha}(-\theta)$. The boxes with the symbol $\tau_1$ represent a free evolution of the system and environment for a period of time $\tau_1$. After the first collision, the resulting quantum state of system and environment is represented by $\rho_t(1)$. $\tau_1$ is the time needed such that a rotation of $\pi/2$ occurs in the system and here it will be given by $\tau_1=1/(4\mathcal{J}_{s, env1})=0.00358$ s.}
\label{nmrcol1}
\end{figure}
\begin{figure}
\includegraphics[scale=1.4]{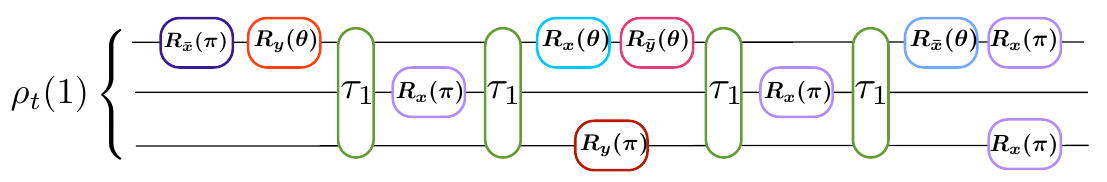}
\caption{Quantum circuit representing the NMR pulse sequence for implementing the second collision. The input state is the resulting state from the first collision $\rho_t(1)$. The boxes with the symbols $R_{\alpha}(\theta)$
indicate that a rotation of the angle $\theta$ on that particular
spin was performed around the $\alpha$ direction and $R_{\bar{\alpha}}(\theta)=R_{\alpha}(-\theta)$. The boxes with the symbol $\tau_1$ represent a free evolution of the system and environment for a period of time $\tau_1$. $\tau_1$ is the time needed such that a rotation of $\pi/2$ occurs in the system and here it will be given by $\tau_1=1/(4\mathcal{J}_{s, env2})=0.00525$ s.}
\label{nmrcol2}
\end{figure}
The NMR implementation of the collisions between the system
and particles of the environment are shown in Figs.~\ref{nmrcol1},\ref{nmrcol2}. The boxes with the symbols $R_{\alpha}(\theta)$
indicate that a rotation of the angle $\theta$ on that particular
spin was performed around the $\alpha$ direction. The big boxes are free evolution of the system for the time $\tau$. 
The angles $\theta$ are related to the parameter $\eta$ of the collisional
model, and they are of the order of 1\textdegree . For the steps used
in the experiment the differences between them were approximately
of 0.5\textdegree . The system we have used is homonuclear, which
means that all of the spins are of the same species (fluorine in this
case), and implementing single rotations were hard because they are
close in frequency. Then, the length of the pulses for exciting only
one spin (qubit) are long. This causes undesired evolutions of the
quantum state due to the interactions between the
spins, leading the whole system to evolve while the individual operations
are applied. Furthermore, the $rf$ pulses are also imperfect and they
may affect others spins, besides the one which it is intended for.
This means that at the end of the quantum algorithm lots of errors
due to these undesirable evolutions and pulse imperfections will have
been accumulated. In order to correct the state of the system the
method described in \cite{key-2} was used. 
\begin{figure}
\includegraphics[scale=1.4]{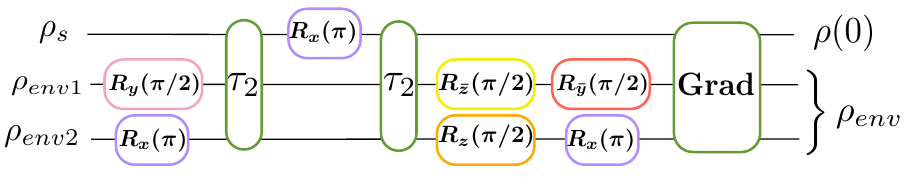}\caption{Quantum circuit representing the NMR pulse sequence for the state preparation of the initial state of system and environment given by $\rho(0)\otimes\rho_{env}$. The boxes with the symbols $R_{\alpha}(\theta)$
indicate that a rotation of the angle $\theta$ on that particular
spin was performed around the $\alpha$ direction and $R_{\bar{\alpha}}(\theta)=R_{\alpha}(-\theta)$. The boxes with the symbol $\tau_2$ represent a free evolution of the system and environment for a period of time $\tau_2$. The box with the symbol Grad stands for the magnetic field gradient that is applied in the $z$-direction.}
\label{statepp-1}
\end{figure}

\begin{figure}
\includegraphics[scale=0.3]{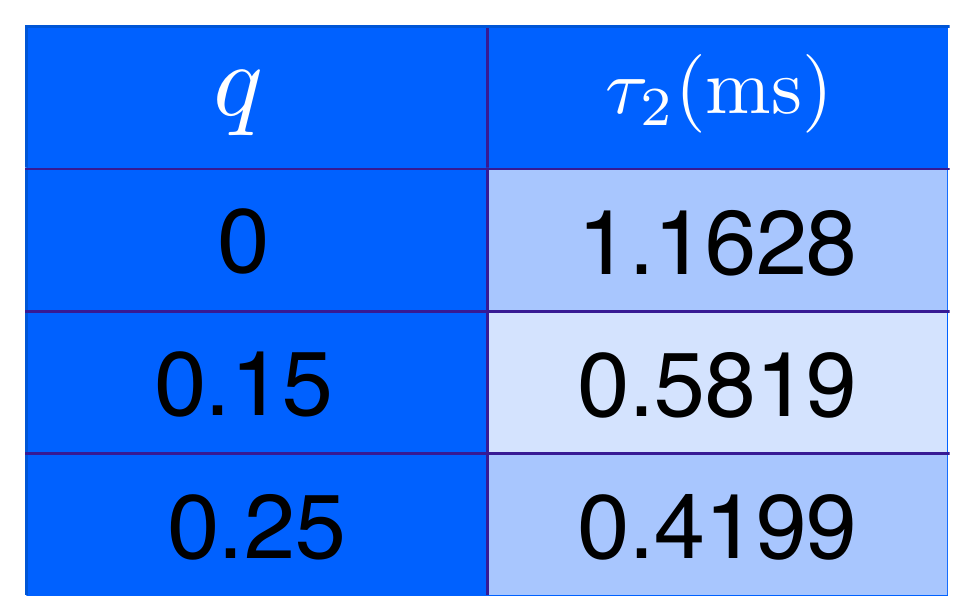}\caption{Parameter $\tau_2$ has a strictly relation to the correlation parameter $q$. The values used in the experiment are presented here.}
\label{table2}
\end{figure}

For the experiment, the values of $\eta$ were very small and this
created a huge source of errors in the operations, since they had
to be performed in tiny steps and, therefore, were not so different.
In fact, the rotations angles were so small that the controlled operations,
which simulate the collisions, were very hard to be performed properly.
In order to achieve the necessary precision in the experiment the
rotations of $\theta$ had to be precisely adjusted for each value
of $\eta$ of the specific collision. For adjusting the angles values,
two parameters may be varied, the amplitude and duration of the radio
frequency pulses. However, varying these two parameters the sequence
of operations needed for the corrections changes as well. Therefore,
the correct pulse sequences were determined by combining the two analysis
and testing in the spectrometer. A good precision could be achieved
then and the small rotations, of the order of 0.30 degrees, could
be implemented. For the state preparation the pulse sequence
shown in Fig.~\ref{statepp-1} was performed, for more details
see \cite{key-3,key-4}. The time $\tau_2$ of the free evolution here has a strict relation to the correlation parameter $q$ and is presented in Fig.~\ref{table2}. A the end of the circuit a magnetic field gradient in the $z$-direction is applied, which works as a transversal relaxation time killing the off-diagonal terms of the density matrix of system and environment.
Finally, after each run, full state tomography was
performed in the state of the system~\cite{key-1}.

Note that this experiment was done much faster ($\sim 0.03$ s) than the typical relaxation times of the elements of the C$_{2}$F$_{3}$I molecule, which is of the order of 1 s.

\subsection*{Acknowledgments}
We would  like  to  thank  the  support  from  the  Brazilian  agencies  CNPq  and CAPES. J.P.S.P., R.S.S, A.M.S, I.R, and I.S.O would like to thank the support of FAPERJ. M.F.S. would like to thank the support of FAPEMIG, project PPM IV. This work is part of
the INCT-IQ from CNPq.


\begin{thebibliography}{99}

\bibitem{kampen} van Kampen, N. G. \textsl{Stochastic Processes in Physics and Chemistry} (Elsevier, Amsterdam, 2007).

\bibitem{chaitin} Chaitin, G. J. \textsl{Information randomness and incompleteness: papers on algorithmic information theory} (World Scientific, Singapore, 1990).

\bibitem{vasile} Vasile, R., Olivares, S., Paris,  M. G. A.  and Maniscalco, S. Continuous-variable quantum key distribution in non-Markovian channels. \textit{Phys. Rev. A} \textbf{83}, 042321 (2011).

\bibitem{matsuzaki} Matsuzaki, Y., Benjamin, S. C. and Fitzsimons, J. Magnetic field sensing beyond the standard quantum limit under the effect of decoherence. \textit{Phys. Rev. A} \textbf{84}, 012103 (2011).

\bibitem{chin} Chin, A. W., Huelga, S. F. and Plenio, M. B. Quantum Metrology in Non-Markovian Environments. \textit{Phys. Rev. Lett.} \textbf{109}, 233601 (2012).

\bibitem{elsi} Laine, E.-M., Breuer, H.-P. and Piilo, J. Nonlocal memory effects allow perfect teleportation with mixed states. \textit{Sci. Rep.} \textbf{4}, 4620 (2014).

\bibitem{Bogna} Bylicka, B., Chruscinski, D. and Maniscalco, S. Non-Markovianity and reservoir memory of quantum channels: a quantum information theory perspective. \textit{Sci. Rep.} \textbf{4}, 5720 (2014).


\bibitem{rau} Rau, J. Relaxation Phenomena in Spin and Harmonic Oscillator Systems. \textit{Phys. Rev.} \textbf{129}, 1880 (1963).

\bibitem{ziman1} Ziman M. and Buzek  V. All (qubit) decoherences: Complete characterization and physical implementation. \textit{Phys. Rev. A} \textbf{72}, 022110 (2005).

\bibitem{ziman2} Ziman, M., Stelmachovic, P. and Buzek, V. Description of Quantum Dynamics of Open Systems Based on Collision-Like Models. \textit{Open Syst. Inf. Dyn.} \textbf{12}, 81 (2005).

\bibitem{giovannetti} Giovannetti, V. and Palma, G. M. Master Equations for Correlated Quantum Channels. \textit{Phys. Rev. Lett.} \textbf{108}, 040401 (2012).

\bibitem{tomas} Ryb\'ar, T., Filipov, S. N. , Ziman,  M. and Buzek, V. Simulation of indivisible qubit channels in collision models. \textit{J. Phys. B: At. Mol. Opt. Phys.}  \textbf{45}, 154006 (2012).

\bibitem{ciccarello} Cicarello, F., Palma, G. M. and Giovannetti, V. Collision-model-based approach to non-Markovian quantum dynamics. \textit{Phys. Rev. A} \textbf{87}, 040103(R) (2013).

\bibitem{vacchini} Vacchini, B. Non-Markovian master equations from piecewise dynamics. \textit{Phys. Rev. A} \textbf{87}, 030101(R) (2013).

\bibitem{budini} Budini, A. A. Embedding non-Markovian quantum collisional models into bipartite Markovian dynamics. \textit{Phys. Rev. A} \textbf{88}, 032115 (2013).

\bibitem{Paternostro} McCloskey, R. and Paternostro, M. Non-Markovianity and system-environment correlations in a microscopic collision model. \textit{Phys. Rev. A} \textbf{89}, 052120 (2014).

\bibitem{nadja1} Bernardes, N. K.,  Carvalho, A. R. R., Monken,  C. H., Santos, M. F. Environmental correlations and Markovian to non-Markovian transitions in collisional models. \textit{Phys. Rev. A} \textbf{90}, 032111 (2014).

\bibitem{Liu} Liu, B.-H. \textit{et al}. Experimental control of the transition from Markovian to non-Markovian dynamics of open quantum systems. Experimental control of the transition from Markovian to non-Markovian dynamics of open quantum systems. \textit{Nat. Phys.} \textbf{7}, 931 (2011).

\bibitem{Liu2} Liu, B.-H. \textit{et al}. Photonic realization of nonlocal memory effects and non-Markovian quantum probes. \textit{Sci. Rep.} \textbf{3}, 1781 (2013).

\bibitem{tang} Tang, J.-S. T \textit{et al}. Measuring non-Markovianity of processes with controllable system-environment interaction. \textit{Europ. Phys. Lett.} \textbf{97}, 10002 (2012).

\bibitem{Steve1} Fanchini,  F., \textit{et al}. Non-Markovianity through accessible information. \textit{Phys. Rev. Lett.} \textbf{112}, 210402 (2014).

\bibitem{Fabio2} Jin, J., \textit{et al}. All-optical non-Markovian stroboscopic quantum simulator. \textit{Phys. Rev. A }\textbf{91}, 012122 (2015).

\bibitem{chiuri} Chiuri, A., Greganti, C., Mazzola, L., Paternostro, M. and Mataloni, P. Linear Optics Simulation of Quantum Non-Markovian Dynamics. \textit{Sci. Rep.} \textbf{2}, 968 (2012).

\bibitem{Xu} Xu, J.-S., \textit{et al}. Experimental recovery of quantum correlations in absence of system-environment back-action. \textit{Nat. Comm.} \textbf{4}, 2851 (2013).

\bibitem{Fabio1} Orieux,  A., \textit{et al}. Experimental on-demand recovery of entanglement by local operations within non-Markovian dynamics. \textit{Sci. Rep.} \textbf{5}, 8575 (2015).

\bibitem{nadja2}  Bernardes, N. K., Cuevas, A. , Orieux, A., Monken, C. H., Mataloni, P., Sciarrino, F.  and Santos,  M. F. Experimental observation of weak non-Markovianity. \textit{Sci. Rep.} \textbf{5}, 17520 (2015).

\bibitem{t1} Batalhao, T. B., Souza, A. M., Sarthour, R. S., Oliveira, I. S., Paternostro, M., Lutz, E. and Serra, R. M. Irreversibility and the Arrow of Time in a Quenched Quantum System. \textit{Phys. Rev. Lett.} \textbf{115}, 190601 (2015).


\bibitem{t2} Auccaise, R., Araujo-Ferreira, A. G., Sarthour, R. S., Oliveira, I. S., Bonagamba, T. J. and Roditi, I. Spin Squeezing in a Quadrupolar Nuclei NMR System. \textit{Phys. Rev. Lett.} \textbf{114}, 043604 (2015).


\bibitem{t3} Batalhao, T. B., Souza, A.  M., Mazzola, L., Auccaise,  R., Sarthour, R. S., Oliveira, I. S., Goold, J., De Chiara, G., Paternostro, M. and Serra, R. M. Experimental Reconstruction of Work Distribution and Study of Fluctuation Relations in a Closed Quantum System. \textit{Phys. Rev. Lett.} \textbf{113}, 140601 ( 2014).
 

\bibitem{t4} Anvari Vind, F. , Foerster, A., Oliveira, I. S., Sarthour, R. S., Soares-Pinto, D. O., Souza A. M. and Roditi, I. Experimental realization of the Yang-Baxter Equation via NMR interferometry. \textit{Sci. Rep.} \textbf{6}, 20789 (2016).

\bibitem{BLP} Breuer, H.-P., Laine, E.-M. and Piilo, J. Measure for the Degree of Non-Markovian Behavior of Quantum Processes in Open Systems. \textit{Phys. Rev. Lett.} \textbf{103}, 210401 (2009).

\bibitem{key-1}  Leskowitz G. M. and Mueller, L. J. State interrogation in nuclear magnetic resonance quantum-information processing. \textit{Phys Rev. A} \textbf{69}, 052302 (2004).

\bibitem{key-2} Ryan, C. A., Negrevergne, C., Laforest, M., Knill, E. and Laflamme, R. Liquid-state nuclear magnetic resonance as a testbed for developing quantum control methods. \textit{Phys. Rev. A} \textbf{78}, 012328 (2008).

\bibitem{key-3} Gershenfeld N. A. and Chuang I. L. Bulk Spin-Resonance Quantum Computation. \textit{Science} \textbf{275}, 350 (1997).

\bibitem{key-4}Cory, D. G., Fahmy, A. F. and Havel, T. F. Ensemble quantum computing by NMR?spectroscopy. \textit{Proc. Natl. Acad. Sci. USA} \textbf{94}, 1634 (1997).

\bibitem{simone} Mukherjee, V., Giovannetti, V., Fazio, R., Huelga, S. F., Calarco, T. and  Montangero, S.  Efficiency of quantum controlled non-Markovian thermalization. \textit{New J. Phys.} \textbf{17} 063031, (2015). 

\bibitem{cesar} Raitz, C., Souza, A. M., Auccaise, R., Sarthour, R. S. and Oliveira, I. S. Experimental implementation of a nonthermalizing quantum thermometer. \textit{Quantum Inf. Process} \textbf{14}, 37 (2015).


\end{thebibliography}

\appendix


\end{document}